\documentclass[aps,preprint,nofootinbib,superscriptaddress,prc]{revtex4-1}
\usepackage{amssymb}

\usepackage{epsfig}
\usepackage[bookmarksnumbered,bookmarksopen,colorlinks,citecolor=blue,linkcolor=blue]{hyperref}

\usepackage{fancyhdr}
\begin{document}

\fancyhead[c]{\small To be published in `Chinese Physics C'} \fancyfoot[C]{\small \thepage}

\title{Fragment Distribution in Reactions of $^{78,86}$Kr+$^{181}$Ta\thanks{Supported by Youth Research Foundation of Shanxi Datong University under Grant No. 2016Q10. }}

\author{Donghong Zhang}

 \affiliation{College of Physics and Electronics, Institute of Theoretical Physics, Shanxi Datong University, Datong 037009, China}
 \email{zhdhsuccess@163.com}

\author{Fengshou Zhang}
\affiliation{The Key Laboratory of Beam Technology and Material Modification of Ministry of Education,College of Nuclear Science and Technology, Beijing Normal University, Beijing 100875, China}
\affiliation{Beijing Radiation Center, Beijing 100875, China}
\affiliation{Center of Theoretical Nuclear Physics, National Laboratory of Heavy Ion Accelerator of Lanzhou, Lanzhou 730000, China}

\begin{abstract}
Within the framework of the isospin-dependent quantum molecular dynamics model along with the GEMINI model, the reaction of $^{86}$Kr+$^{181}$Ta at 80,120 and 160 MeV/nucleon and the reaction of $^{78}$Kr+$^{181}$Ta at 160 MeV/nucleon are studied, and the production cross sections of the generated fragments are calculated. More intermediate and large mass fragments can be produced in the reaction with a large range of impact parameter. The production cross sections of nuclei such as the isotopes of Si and P generally decrease with the increasing incident energy. The isotopes near the neutron drip line are produced more in the neutron-rich system $^{86}$Kr+$^{181}$Ta.
\end{abstract}

\maketitle

\begin{center}
\textbf{I. INTRODUCTION}
\end{center}

The research of new nuclide is an important subject in the field of nuclear physics\cite{lab1}. With the emergence of the powerful detectors, the general characteristics of multifragmentation have been studied\cite{lab2,lab3,lab4,lab5,lab6}. Further development in the future will be related to the study of many observations and the correlation of multifragmentation events. As an effective way to produce rare isotopes, the nuclear multifragmentation plays an important role in the study of nuclear physics\cite{lab7}.

The stable nuclide are located in the narrow region of the nuclide map, and the line that runs through the center of the region is called $\beta$ stability line. The theoretical models of structures, such as the shell model, the liquid drop model, the collective model\cite{lab8}, are based on the study of the nuclei located in the stability line and nearby. With the development of nuclear physics and the progress of accelerator and nuclear detection technology, many new nuclide have been synthesized by nuclear reaction\cite{lab9,lab10}. The nuclei on the nuclide map have been expanded in the direction of both proton number and neutron number.

In recent years, more and more attention has been paid to the experimental and theoretical research of the exotic nuclei far away from the $\beta$ stability line\cite{lab11,lab12}. The area of the nuclide map near the drip line has been widely concerned\cite{lab13,lab14}, which is very important for explaining the change of nuclear structure with the increase of neutron-proton ratio and the study of the mechanism of nucleosynthesis\cite{lab15,lab16,lab17,lab18}. Therefore, it is of great significance to study the production of the isotopes near the drip line.

This article is based on the isospin-dependent quantum molecular dynamics (IQMD) model along with the statistical decay model GEMINI to study the production cross sections of nuclide in heavy ion collisions. By investigating the reactions of different collision systems, the multiplicity, charge distribution and production cross sections of the nuclide near the drip line are calculated, and the production cross sections of the isotopes of Si and P are obtained. The results show that the production cross sections of isotopes in the reaction are related to the incident energy and the isospin of the collision system. The production cross sections of the isotopes of Si and P decrease with the increasing energy, and the isotopes near the neutron drip line are more productive in the reaction of $^{86}$Kr+$^{181}$Ta than in the reaction of $^{78}$Kr+$^{181}$Ta.

\begin{center}
\textbf{II. THEORETICAL FRAMEWORK}
\end{center}

Since the fragments are produced in the kinetic reaction, it is necessary to develop micro-kinetic models for studying the formation of fragments\cite{lab19,lab20,lab21}. Some of the existing models are based on statistical descriptions of multi-body phase space calculations\cite{lab22,lab23,lab24,lab25,lab26} and others are molecular dynamics models\cite{lab27,lab28,lab29} or stochastic mean field models\cite{lab30,lab31} that describe the dynamical evolution of the system in the nuclear collision. The first method uses the equilibrium state statistical mechanics method to study the thermodynamic description of finite nuclear systems. The second method is a complete description of the temporal evolution of the collision system and is therefore useful for studying nuclear species, finite-size effects, kinetics of phase transitions and so on. The empirical parameterization of fragment cross sections can help to predict the mass and charge distribution of heavy ion reactions. Statistical models can reproduce the experimental results of heavy ion collisions. The molecular dynamics model includes information about the transport mechanism. The micro-antisymmetric molecular dynamics model\cite{lab32} and the fermionic molecular dynamics model\cite{lab33} have been developed. The isospin-dependent Boltzmann-Langevin equation (IBLE) model\cite{lab34} can also be used to calculate the cross section of fragments. The quantum molecular dynamics (QMD) model and statistical decay model GEMINI are used to describe heavy ion reactions.

The IQMD model\cite{lab35} is a model considering isospin freedom on the basis of QMD, which contains the isospin degree of freedom of the nucleons. The IQMD model can be well applied to the study of many heavy ion collisions at intermediate energy. As a multi-body theory for simulating heavy ion reactions with incident energies between 30 MeV/nucleon and 1 GeV/nucleon, the IQMD model uses Gauss wave packet to describe every nucleon
\begin{equation}
  \phi_i(\mathbf{r},t)=\frac{1}{({2{\pi}L})^{3/4}}e^{{-}\frac{[\mathbf{r}-\mathbf{r}_i(t)]^2}{4L}}e^{i{\frac{
  \mathbf{r}{\cdot}\mathbf{p}_i(t)}{\hbar}}},
 \end{equation}
where $\mathbf{r}_{i}$ and $\mathbf{p}_{i}$ represent the center of the coordinate space and the momentum space of the $i$th nucleon, and $L$ represents the corresponding wave packet width. The N body wave function can be represented by the direct product of the coherent states:
\begin{equation}
    \Phi(\textbf{r},\textbf{r}_{1},\cdots,\textbf{r}_{N},\textbf{p}_{1},\cdots,\textbf{p}_{N},t) =
     \prod_{i} \phi_{i}(\textbf{r},\textbf{r}_{i},\textbf{p}_{i},t).
    \label{phit}
 \end{equation}
The antisymmetry is not considered here. The values of the initial parameter adopted can make the density distribution and momentum distribution of all the nuclei of the projectile and target have the proper distribution. The evolution of the system is derived from the generalized variational principle\cite{lab36}:
\begin{equation}
     S = \int _{t_{1}} ^{t_{2}} \textbf{L}[\Phi,\Phi^{*}]dt
     \label{variat},
 \end{equation}
where \textbf{L} is Lagrange function:
  \begin{equation}
     \textbf{L} = \langle \Phi | i\hbar\frac{d}{dt}-H | \Phi \rangle
     \label{langr}.
 \end{equation}
Derivation of time here includes the derivation of the parameters $\textbf{r}_{i}$ and $\textbf{p}_{i}$. By taking variation on the action S, the evolution of the parameters $\textbf{r}_{i}$ and $\textbf{p}_{i}$ over time can be described by the Euler-Lagrange equation:

\begin{equation}
    $$
     $\frac{d}{dt} \frac{\partial \textbf{L}}{\partial \textbf{\.{p}}_{i}}$
       -$\frac{\partial \textbf{L}}{\partial \textbf{p}_{i}} = 0  \qquad \rightarrow$
       $\qquad \textbf{\.{r}}_{i}=\frac{\partial \langle H \rangle}{\partial \textbf{p}_{i}}, $\\
       $\frac{d}{dt} \frac{\partial \textbf{L}}{\partial \textbf{\.{r}}_{i}}$
       -$\frac{\partial \textbf{L}}{\partial \textbf{r}_{i}} = 0  \qquad \rightarrow$
       $\qquad \textbf{\.{p}}_{i}=\frac{\partial \langle H \rangle}{\partial \textbf{r}_{i}},$

 $$
 \end{equation}
 
Based on Wigner transform on the wave function, N body phase space distribution function can be expressed as:
\begin{equation}
  f(\mathbf{r},\mathbf{p},t)=\sum_{i=1}^{n}\frac{1}{({\pi}{\hbar})^3}e^{{-}\frac{[\mathbf{r}-\mathbf{r}_i(t)]^2}{2L}}e^{{-}\frac{[\mathbf{p}-\mathbf{p}_i(t)]^2{\cdot}{2L}}{{\hbar}^2
  }}.
 \end{equation}
The evolution of the nuclei in the mean field over time in the system can be described by the Hamiltonian equation of motion:
\begin{equation}
  {\mathbf{\dot{r}}}_{i}=\nabla_{\mathbf{p_{i}}}H,{\mathbf{\dot{p}}}_{i}=-\nabla_{\mathbf{r_{i}}}H.
 \end{equation}

The statistical model GEMINI\cite{lab37} can well describe the series of decay of the thermonuclear system. All decay chains adopt the Monte-Carlo method until the resulting {products can not decay further. The decay width can be calculated from the light-particle evaporation formula of Hauser-Feshbach\cite{lab38} and the symmetric splitting transition formula of Moretto.

\begin{center}
\textbf{III. RESULTS AND DISCUSSION}
\end{center}

The production cross section of Fe in the reaction of $^{86}$Kr+$^{181}$Ta at 64 MeV/nucleon is depicted in Fig.~\ref{fig1}. For comparison, the experimental data and EPAX calculations have also been shown. The simulation results are in good agreement with the experimental data, but EPAX calculations underestimate the experimental ones. The cross sections of fragments are mainly affected by the potential parameters in the model and the selection of collision events.

\begin{figure}\center
\psfig{file=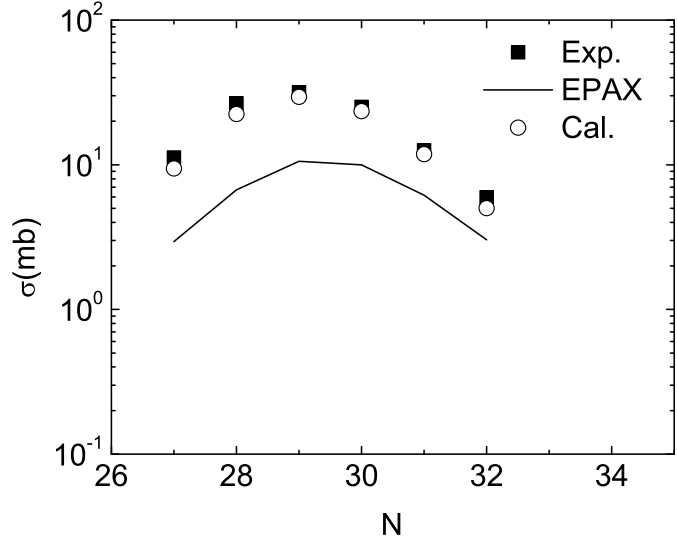,width= 0.8\textwidth}
\caption{The cross sections for the Fe isotopes predicted
in this work (open circles), the experimentally measured\cite{lab39}(solid squares) and EPAX calculations(solid curves) in the reaction of $^{86}$Kr+$^{181}$Ta at 64 MeV/nucleon.}\label{aba:fig1}
\end{figure}

As the reaction conditions of the stimulated system, the impact parameters have an important impact on the reaction mechanism. Fig.~\ref{fig2} shows the charge distribution of the reaction of $^{86}$Kr+$^{181}$Ta at 160 MeV/nucleon under different impact parameters. As can be seen from the figure, given the same minimum value of impact parameter, intermediate and large mass fragments produced more in the reaction with larger value range of impact parameter, while the production cross sections of light mass fragments weakly depend on the value of impact parameter. The large difference between the fragments of large Z are due to the isospin effects in projectile fragmentations. This has been well understood in theory\cite{lab40,lab41,lab42,lab43,lab44}. The isopin difference between the core and skirt of the projectile nucleus influences the difference between the neutron and proton density distribution in these areas, induces the difference of fragments in large impact parameters ranges. The impact parameter used in this work is $b=0-10fm$.

\begin{figure}\center
\psfig{file=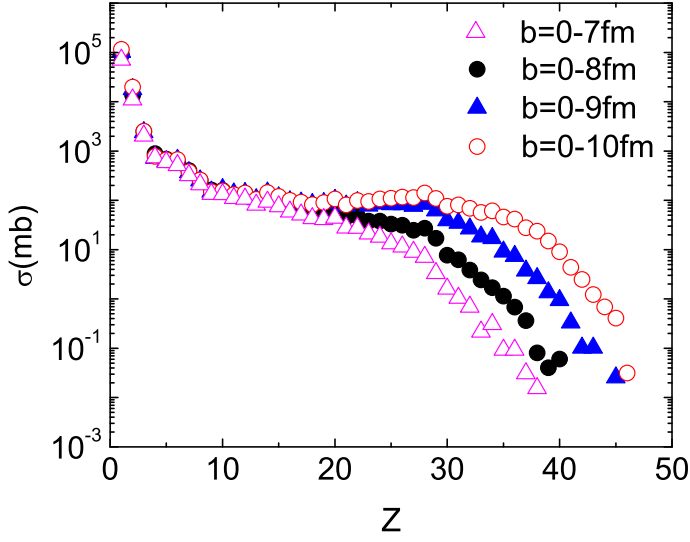,width= 0.8\textwidth}
\caption{The relationship between the charge distribution and the impact parameters in the reaction of $^{86}$Kr+$^{181}$Ta at 160 MeV/nucleon.}\label{aba:fig2}
\end{figure}

In order to investigate the energy and isospin dependence of the charge distributions, the isotopes of Si and P in each reaction system at different incident energies are calculated using IQMD and GEMINI models. Fig.~\ref{fig3} shows the cross section of the S isotopes produced by the reaction of $^{86}$Kr+$^{181}$Ta at incident energies of 80£¬120 and 160 MeV/nucleon. Among them, $^{42}$Si is a new nucleus that has not been synthesized experimentally. The results in the figure show that the peaks of the production cross sections of Si locate at $^{28-30}$Si at different incident energies. In the process of increasing the incident energy from 80 MeV/nucleon to 160 MeV/nucleon, the production cross section of $^{22-42}$Si decreases, and the difference of the production cross section of $^{24}$Si between the incident energies of 80 MeV/nucleon and 160 MeV/nucleon is more obvious.

\begin{figure}\center
\psfig{file=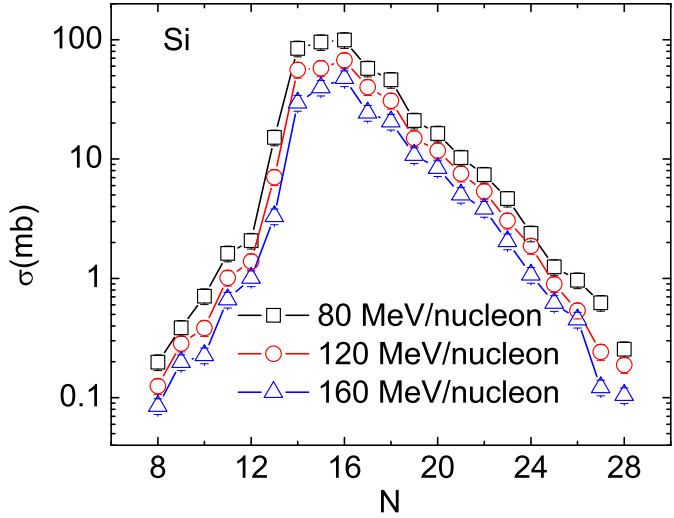,width= 0.8\textwidth}
\caption{The production cross sections of the Si isotopes for the reactions of $^{86}$Kr+$^{181}$Ta at 80 MeV/nucleon, 120 MeV/nucleon and 160 MeV/nucleon.}\label{aba:fig3}
\end{figure}

The production cross sections of the Si isotopes in the reactions of $^{86}$Kr+$^{181}$Ta and $^{78}$Kr+$^{181}$Ta at 160 MeV/nucleon are plotted in Fig.~\ref{fig4}. It can be seen from the figure that $^{41}$Si is not produced in the reaction of $^{78}$Kr+$^{181}$Ta£¬the production cross sections of the isotopes near the proton drip line such as $^{22-26}$Si in the reaction of $^{78}$Kr+$^{181}$Ta are larger than those in $^{86}$Kr+$^{181}$Ta£¬while the production cross sections of the isotopes near the neutron drip line such as $^{35-40}$Si and $^{42}$Si in the reaction of $^{86}$Kr+$^{181}$Ta are larger than those in $^{78}$Kr+$^{181}$Ta. The peak values at $^{28-30}$Si are roughly the same in both reactions.

\begin{figure}\center
\psfig{file=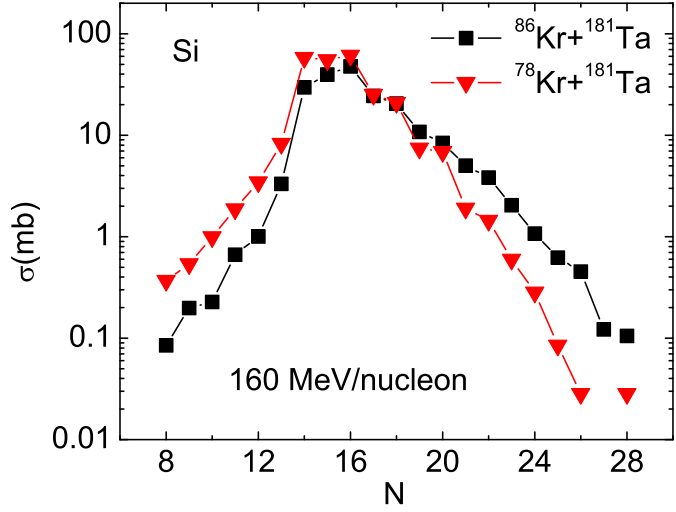,width= 0.8\textwidth}
\caption{The production cross sections of the Si isotopes in the reactions of $^{86}$Kr+$^{181}$Ta and $^{78}$Kr+$^{181}$Ta at 160 MeV/nucleon.}\label{aba:fig4}
\end{figure}

Fig.~\ref{fig5} shows the production cross sections of the P isotopes in the reactions of $^{86}$Kr+$^{181}$Ta at 80-160 MeV/nucleon. Among them, $^{46}$P is an unknown nucleus that has not been synthesized experimentally. The results in the figure show that the peak positions of the production cross section of P locate at $^{31-33}$P at different incident energies. The production cross section of $^{24-46}$P decreases in the process of increasing the incident energy from 80 MeV/nucleon to 160 MeV/nucleon, while the gap of the production cross section of $^{27}$P and $^{45}$P between the incident energies of 80 MeV/nucleon and 160 MeV/nucleon is obvious.

\begin{figure}\center
\psfig{file=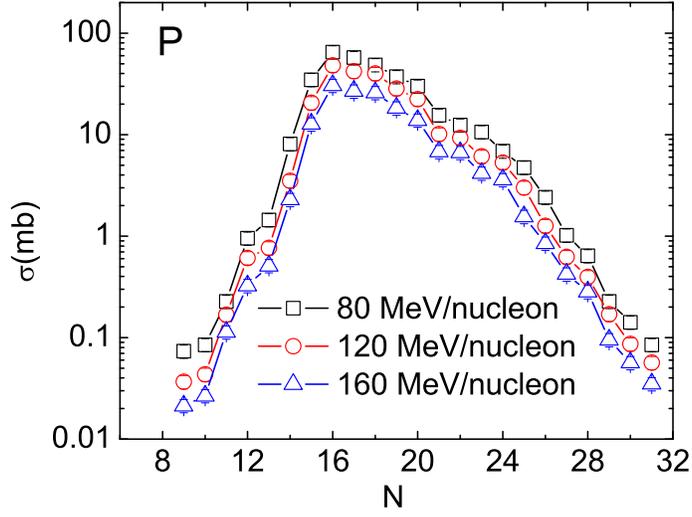,width= 0.8\textwidth}
\caption{The production cross sections of the P isotopes for the reactions of $^{86}$Kr+$^{181}$Ta at 80 MeV/nucleon, 120 MeV/nucleon and 160 MeV/nucleon.}\label{aba:fig5}
\end{figure}

The production cross sections of the P isotopes in the reactions of $^{86}$Kr+$^{181}$Ta and $^{78}$Kr+$^{181}$Ta at 160 MeV/nucleon are depicted in Fig.~\ref{fig6}. It can be noted that the products of $^{44-46}$P are not found in the reaction of $^{78}$Kr+$^{181}$Ta, the production cross sections of the isotopes near the proton drip line such as $^{24-28}$P are larger than those in $^{86}$Kr+$^{181}$Ta, while the isotopes near the neutron drip line such as $^{38-43}$P are more produced in the reaction of $^{86}$Kr+$^{181}$Ta. The peak values locate at $^{31-33}$P and are approximately the same in both reactions.

\begin{figure}\center
\psfig{file=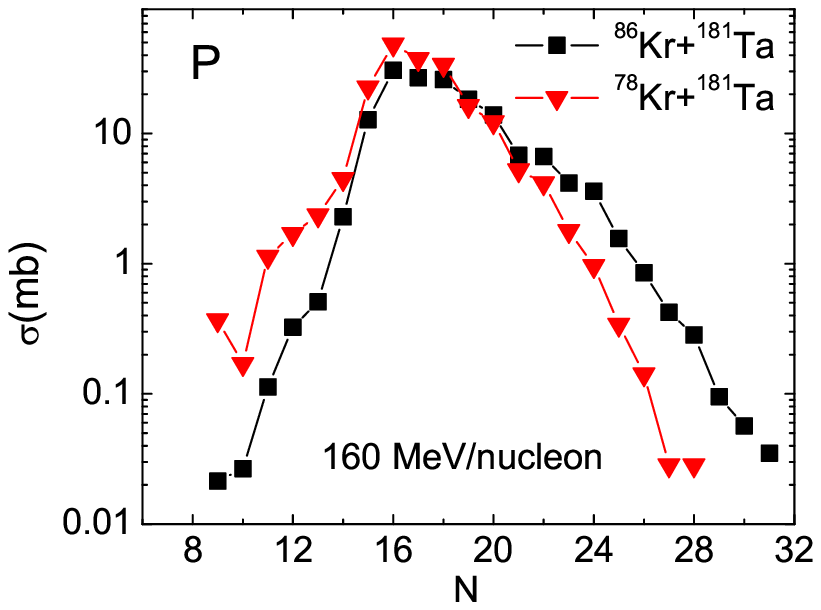,width= 0.8\textwidth}
\caption{The production cross sections of the P isotopes in the reactions of $^{86}$Kr+$^{181}$Ta and $^{78}$Kr+$^{181}$Ta at 160 MeV/nucleon.}\label{aba:fig6}
\end{figure}

As can be seen from the above figures, the intermediate and large mass fragments generate more in the reactions with same minimum value but larger maximum value of impact parameter, while light mass fragments are less affected by the range of impact parameter. This is mainly due to the different reaction mechanisms of system for different impact parameters. The system mainly performs the fusion reaction when the impact parameters are small. As the impact parameters increase, the ratios of the fast fission and the deep inelastic collision increase, resulting in more heavy fragments. The production cross sections of the isotopes of Si and P in the reaction of $^{86}$Kr+$^{181}$Ta generally decrease with the increasing energy at 80 MeV/nucleon to 160 MeV/nucleon. For the same incident energy, the production cross sections of the isotopes near the proton drip line in the reaction of $^{78}$Kr+$^{181}$Ta are larger than those in the reactions of $^{86}$Kr+$^{181}$Ta, while the production cross sections of the isotopes near the neutron drip line in the reaction of $^{86}$Kr+$^{181}$Ta are larger than those in the other reactions. This phenomenon is mainly caused by the isospin effect in the nuclear multifragmentation. Because the reaction conditions are exactly the same except for the neutron-proton ratio. For stable nuclides, the production cross sections in two reactions are very close.

\begin{center}
\textbf{IV. CONCLUSIONS }
\end{center}

The fragment distribution in the reactions of $^{86}$Kr+$^{181}$Ta and $^{78}$Kr+$^{181}$Ta at 80-160 MeV/nucleon are studied via the IQMD model accompanied by the GEMINI model. It is found that the intermediate and large mass fragments can be produced more in the reactions with same minimum value but larger maximum value of impact parameter, while the value of impact parameter has less effect on the light mass fragments. This is mainly due to the different reaction mechanisms of the system for different impact parameters. The channel of the system is the fusion when the impact parameters are small. As the impact parameters increase, the ratio of the fast fission and deep inelastic collision also increases. The production cross sections of the isotopes of Si and P produced in the reaction of $^{86}$Kr+$^{181}$Ta generally decrease with the increasing energy. For the same incident energy, the production cross sections of the isotopes near the proton drip line in the reaction of $^{78}$Kr+$^{181}$Ta are larger than those in the reaction of $^{86}$Kr+$^{181}$Ta, while the production cross sections of the isotopes near the neutron drip line in the neutron-rich system of $^{86}$Kr+$^{181}$Ta are larger than those in the other systems. The phenomenon is mainly caused by the isospin effect of heavy ion reaction. For stable nuclide, the difference of the production cross section between the two reactions is very slight. These results may provide some guidance on how to select the reaction system and incident energy to produce the unknown nuclide and to conduct further relevant investigations.

\vspace{-1mm}
\centerline{\rule{80mm}{0.1pt}}

\end{document}